\documentstyle[preprint,aps]{revtex}

\begin{document}
\draft
\preprint{FNUSAL - 1/95}

\title{Quark Cluster Model Study of Isospin-Two
Dibaryons}

\author{A. Valcarce $^{(1)}$, H. Garcilazo $^{(2)}$
and F. Fern\'andez$^{(1)}$}

\address{$(1)$ Grupo de F\' \i sica Nuclear \\
Universidad de Salamanca, E-37008 Salamanca, Spain}

\address{$(2)$ Escuela Superior de F\' \i sica y Matem\'aticas \\
Instituto Polit\'ecnico Nacional, Edificio 9,
07738 M\'exico D.F., Mexico}

\maketitle

\begin{abstract}

Based on a quark cluster model for the non-strange sector
that reproduces reasonably well the nucleon-nucleon system
and the excitation of the $\Delta$ isobar, we generate a
nucleon-$\Delta$ interaction and present the
predictions for the several
isospin two channels.
The only attractive channels are $0^+$ and $0^-$, but not
attractive enough to generate a resonance. If a resonance is
artificially generated and is required to have the observed
experimental mass, then our model predicts a width that agrees
with the experimental result.
\end{abstract}

\narrowtext
\newpage

\section{ Introduction}

The possibility that a resonance may exist in the $\pi NN$
system with $J^P = 0^-$ has been
the subject of great discussion recently \cite{BIL1,GAR1,BIL2}.
A $0^-$
resonance in the isospin-0 channel was proposed in Ref. \cite{BIL1} in
order to explain a peculiar behavior of the pionic double charge exchange
cross section in several nuclei from $^{14}C$ to $^{48}Ca$. However, it was
pointed out in Ref. \cite{GAR1} that according
to the calculations of Ref. \cite{GAR2},
a $0^-$ resonance with isospin zero is not possible and that
most likely the resonance must have isospin two. The motivation of this
work is to present the predictions of a quark-model based
interaction about possible resonances in the isospin two sector.

While the isospin one channel has been extensively studied,
the isospin two is less well-known. However, there have
been several works in the literature where the possibility
of a bound state or a resonance in the $\pi NN$ system with
isospin two has been studied. Gale and Duck \cite{GAL}
studied this problem in the framework of the non-relativistic
Faddeev equations, using a model of a rank-one separable potential
for the $\pi$-nucleon interaction in the $P_{33}$ channel without
including the important nucleon-nucleon interaction. Their study
did not support the existence of $\pi NN$ bound states, not even
when they included relativistic kinematics in the Faddeev equations.
They predicted a broad resonance at an energy
about 350 MeV above the $NN$ threshold. In subsequent works by
Ueda \cite{UED} and Kalbermann and Eisenberg \cite{KAL} the
system was investigated using the Heitler-London-Pauli
variational method in a nonrelativistic approach. These authors,
like in Ref. \cite{GAL}, did not find such bound states.
However, the possible existence of a quasibound system was
not ruled out.

Lately, there were a series of papers by
Garcilazo et al. \cite{GAR1,GAR2,GAR3} based on the relativistic
Faddeev equations including both $\pi$-nucleon and nucleon-nucleon
interactions. They used separable potentials for the $\pi$-nucleon
and nucleon-nucleon subsystems with $J \le 2$ which were fitted
to the phase shifts up to 350 MeV.
They found that bound state solutions are possible
when a sufficiently short-ranged $\pi$-nucleon interaction
is assumed. At the same time, they showed that the most likely
configuration to have a bound state or a resonance is that with
angular momentum and parity quantum numbers $J^P = 2^-$.

There have been other calculations about the same problem. Arenh\"ovel
\cite{ARE} predicted a bound state
with the quantum numbers $2^+$ and isospin two,
making use of a nucleon-$\Delta$ interaction mediated by the
exchange of the $\pi$ and $\rho$ mesons (see also ref. \cite{GART}).

Our aim in this work is to present the predictions based on  a
quark dynamical model. This model is able to describe correctly
the nucleon-nucleon system as well as the deuteron. As it is
based only on interactions between quarks its generalization to the
nucleon-$\Delta$ system is straightforward and it does not mean
the inclusion of any new parameter. Therefore, it is our goal to use
the nucleon-$\Delta$ interaction generated by this quark model to
analyze the possibility of the existence of a bound state or a
resonance in the isospin-2 sector in all the
nucleon-$\Delta$ channels with
angular momentum $J \le 2$.

In Section II we will briefly describe the quark model. In Section III
we will review our formalism. In Section IV we will present and
discuss our results. Finally, in Section V we will summarize our
conclusions.

\section{The model of interaction}

The constituent quark model for the nucleon-nucleon ($NN$) interaction
developed during the last decade \cite{FAE}
has been recently extended in a
serie of papers \cite{FVJ,VAL} by introducing a scalar meson
(the $\sigma$-meson) exchange
between quarks and not between the nucleons as done
previously. In this work, the $\sigma$-meson parameters
are related to those of the pion exchange and to the
constituent quark mass through
chiral symmetry requirements.
This model has been successfully applied to the description of
the $NN$ scattering phase
shifts \cite{FVJ} and the static and
electromagnetic properties of the deuteron \cite{VAL}.
At the same time it has been used to the description of processes that
take place with the excitation of a $\Delta$ isobar \cite{FNU} and to study
the pure nucleon-$\Delta$ interaction \cite{ANU}.

The model has been introduced in Refs. \cite{FVJ,VAL}.
The ingredients of the quark-quark
interaction are the confining potential (CON),
the one-gluon exchange (OGE), the
one-pion exchange (OPE) and the one-sigma exchange (OSE).
The explicit form of these
interactions is given by
(See Refs. \cite{FVJ,VAL} for details.),

\begin{eqnarray}
V_{CON} ({\vec r}_{ij}) & = &
-a_c \, {\vec \lambda}_i \cdot {\vec
\lambda}_j \, r^2_{ij} \,  , \\
V_{OGE} ({\vec r}_{ij}) & = &
{1 \over 4} \, \alpha_s \, {\vec
\lambda}_i \cdot {\vec \lambda}_j
\Biggl \lbrace {1 \over r_{ij}} -
{\pi \over m^2_q} \, \biggl [ 1 + {2 \over 3}
{\vec \sigma}_i \cdot {\vec
\sigma}_j \biggr ] \, \delta({\vec r}_{ij})
- {3 \over {4 m^2_q \, r^3_{ij}}}
\, S_{ij} \Biggr \rbrace \, , \\
V_{OPE} ({\vec r}_{ij}) & = & {1 \over 3}
\, \alpha_{ch} {\Lambda^2  \over \Lambda^2 -
m_\pi^2} \, m_\pi \, \Biggr\{ \left[ \,
Y (m_\pi \, r_{ij}) - { \Lambda^3
\over m_{\pi}^3} \, Y (\Lambda \,
r_{ij}) \right] {\vec \sigma}_i \cdot
{\vec \sigma}_j + \nonumber \\
 & & \left[ H( m_\pi \, r_{ij}) - {
\Lambda^3 \over m_\pi^3} \, H( \Lambda \,
r_{ij}) \right] S_{ij} \Biggr\} \,
{\vec \tau}_i \cdot {\vec \tau}_j \, , \\
V_{OSE} ({\vec r}_{ij}) & = & - \alpha_{ch} \,
{4 \, m_q^2 \over m_{\pi}^2}
{\Lambda^2 \over \Lambda^2 - m_{\sigma}^2}
\, m_{\sigma} \, \left[
Y (m_{\sigma} \, r_{ij})-
{\Lambda \over {m_{\sigma}}} \,
Y (\Lambda \, r_{ij}) \right] \, .
\end{eqnarray}

The main advantage of this model comes from the fact that
it works with a single $qq$-meson vertex. Therefore,
its parameters (coupling constants, cut-off masses,...) are
independent of the baryon to which the quarks are coupled, the
difference among them being generated by SU(2) scaling. This
makes the generalization of the $NN$ interaction to any other
non-strange baryonic system straightforward.
Moreover, as explained
in Refs. \cite{FVJ,VAL} the parameters associated with the
scalar field are related to those of the pion
exchange and the constituent quark mass
through: $g_{\sigma NN}^2 / 4\pi = 36 \, \alpha_{ch} m^2_q/m^2_{\pi}$,
$\Lambda_{\sigma}=\Lambda_\pi$ and
$m_{\sigma}^2 = (2 m_q)^2 + m_{\pi}^2$.

Once the quark-quark interaction is chosen, an effective
nucleon-$\Delta$ potential can be obtained in the Born-Oppenheimer
approximation, as the expectation value of the energy of the
six-quark system minus the self-energies of the two clusters,
which can be computed as the energy of the six-quark system
when the two quark clusters do not interact

\begin{equation}
V_{N \Delta (L \, S \, T) \rightarrow N \Delta (L' \, S' \, T)} (R) =
\xi_{L \,S \, T}^{L' \, S' \, T} (R) \, - \,
\xi_{L \,S \, T}^{L' \, S' \, T} (\infty) \, ,
\label{Poten1}
\end{equation}

\noindent
where

\begin{equation}
\xi_{L \, S \, T}^{L' \, S' \, T} (R) \, = \,
{{\left \langle \Psi_{N \Delta}^{L' \, S' \, T} (R) \mid
\sum_{i<j=1}^{6} V_{qq}({\bf r}_{ij})
\mid \Psi_{N \Delta}^{L \, S \, T} (R) \right \rangle}
\over
{\sqrt{\left \langle \Psi_{N \Delta}^{L' \, S' \, T} (R) \mid
\Psi_{N \Delta}^{L' \, S' \, T} (R) \right \rangle}
\sqrt{\left \langle \Psi_{N \Delta}^{L \, S \, T} (R) \mid
\Psi_{N \Delta}^{L \, S \, T} (R) \right \rangle}}} \, .
\label{Poten2}
\end{equation}

\noindent
The wave function is discussed in detail in Refs. \cite{FNU,ANU}.
In expression\ (\ref{Poten2}) the
quark coordinates are integrated out
keeping R fixed, the resulting interaction $V_{N \Delta}$ being
a function of the nucleon-$\Delta$ relative distance.
We have neglected the non-local terms of the potential,
however, for the qualitative analysis we are interested in, the
Born-Oppenheimer approximation is known to provide
good results \cite{new}.
The parameters of the model are summarized in Table I.

\section{Results}

In order to determine the nature (attractive or repulsive)
of a given nucleon-$\Delta$ ($N\Delta$) channel,
we will first calculate the Fredholm determinant of that channel
as a function of energy assuming a stable
delta and nonrelativistic kinematics. That means, we will use the
Lippmann-Schwinger equation

\begin{equation}
T_{ij}(q,q_0)=V_{ij}(q,q_0)+\sum_k \int_0^\infty q^{\prime 2} dq'
V_{ik}(q,q')G_0(E,q')T_{kj}(q',q_0) \, ,
\end{equation}

\noindent
where the two-body propagator is

\begin{equation}
G_0(E,q)={1 \over E-q^2/2\eta+i\epsilon} \, ,
\label{prop}
\end{equation}

\noindent
with reduced mass

\begin{equation}
\eta={m_Nm_\Delta \over m_N+m_\Delta} \, .
\end{equation}

\noindent

The energy and on-shell momentum are related as

\begin{equation}
E=q_0^2/2\eta \, ,
\end{equation}

\noindent
and we will restrict ourselves to the region $E \le 0$.

If we replace the integration in Eq. (7) by a numerical quadrature, the
integral equations take the form

\begin{equation}
T_{ij}(q_n,q_0)=V_{ij}(q_n,q_0)+\sum_k \sum_m w_m q^2_m
V_{ik}(q_n,q_m)G_0(E,q_m)T_{kj}(q_m,q_0) \, ,
\end{equation}

\noindent
where $q_m$ and $w_m$ are the abscissas
and weights of the quadrature (we
use a 40-point Gauss quadrature \cite{abra}).
Eq. (11) gives rise to the
set of inhomogeneous linear equations

\begin{equation}
\sum_k \sum_m M^{ik}_{nm}(E)T_{kj}(q_m,q_0)=V_{ij}(q_n,q_0) \, ,
\end{equation}
with

\begin{equation}
M_{nm}^{ik}(E)=\delta_{ik}\delta_{nm}- w_m q^2_m
V_{ik}(q_n,q_m)G_0(E,q_m) \, .
\end{equation}
If a bound state exists at an energy $E_B$,
the determinant of the matrix
$M_{nm}^{ik}(E_B)$ (the Fredholm determinant)
must vanish, i.e.,

\begin{equation}
\left|M_{nm}^{ik}(E_B)\right|=0.
\end{equation}
Even if there is no bound state, the Fredholm
determinant is a very useful
tool to determine the nature of a given channel.
If the Fredholm determinant is
larger than one that means that channel is repulsive. If the Fredholm
determinant is less than one that means the channel is attractive.
Finally, if the Fredholm determinant passes through zero that means
there is a $N\Delta$ bound state at that energy.
We show in Fig. 1 the Fredholm determinant of the $N\Delta$ channels
with $J\le 2$ and parity + or $-$ as functions of energy
(the channel $2^+$ is not shown since it is overwhelmingly repulsive).
As seen from this
figure, only the channels $0^+$ and $0^-$ are attractive.

Previous calculations of the isospin-2 channels have been based in the
three-body formalism of the $\pi NN$ system \cite{GAR2,GAL}.
In these works \cite{GAR2,GAL}, the possibility of resonances
was investigated by calculating the Fredholm determinant of the various
channels for energies below the $\pi NN$ threshold. We show in Table II the
comparison of our results for the nature of the six states with those
obtained in Refs. \cite{GAR2} and \cite{GAL}.
The difference observed between the predictions of the three-body
models is due to the fact that in Ref. \cite{GAL} the nucleon-nucleon
interaction was neglected.
As seen from this table, the results of
the three-body calculations and our calculation
agree with each other in the
case of the $0^-$, $1^+$, and $2^+$ channels.
The three-body models predict a resonance
in the $2^-$ channel while our result is that this channel is
repulsive. Both three-body calculations predict a large attraction in the
$0^-$ channel which is in agreement with our model
since we have found that this channel
is the most attractive one. In order to
determine whether a resonance exists in
either the $0^-$ or $0^+$
channels, it is necessary to calculate the $N\Delta$
amplitudes for $E>0$.

Since we are looking for possible resonances which decay
into a pion and two nucleons, only those channels which are attractive or
have a $N\Delta$ bound state are good resonance candidates.
In order to investigate whether a given attractive channel possesses a
resonance, we will calculate Argand diagrams
between a stable and an unstable particle for the various
$N\Delta$ states, using the formalism
of ref. \cite{GART}.
In this case, however, we will use relativistic kinematics
and will include the width of the delta. That means, instead of the
propagator (\ref{prop}) we will use \cite{GART}

\begin{equation}
G_0(S,q)={2m_\Delta \over s-m_\Delta^2+im_\Delta\Gamma_\Delta(s,q)} \, ,
\end{equation}

\noindent
where $S$ is the invariant mass squared of the system,
while $s$ is the
invariant mass squared of the $\pi N$ subsystem (those
are the decay products
of the $\Delta$) and is given by

\begin{equation}
s=S+m_N^2-2\sqrt{S(m_N^2+q^2)} \, .
\end{equation}

The width of the $\Delta$ is taken to be \cite{GART}

\begin{equation}
\Gamma_\Delta(s,q)={2 \over 3}\,0.35\, p_0^3 {\sqrt{m_N^2+q^2}\over
m_\pi^2\sqrt{s}} \, ,
\end{equation}

\noindent
where $p_0$ is the pion-nucleon relative momentum given by

\begin{equation}
p_0=\left([s-(m_N+m_\pi)^2][s-(m_N-m_\pi)^2] \over 4s \right)^{1/2} \, .
\end{equation}

We show in Figs. 2 and 3 the phase shifts of the two attractive
channels $0^+$ and $0^-$. As it can be seen from these figures, the
attraction in both channels is not strong enough to produce a
resonance (it does not reach 90 degrees).
As an example of the predictions of our model for other channels,
we show in Fig. 4
the Argand diagram corresponding to the $^3P_2$, $^5P_2$
and $^3F_2$
amplitudes of the $2^-$ channel. The $^5F_2$ amplitude
is not shown since it is much smaller. As seen in this
figure, the $^5P_2$ wave is strongly repulsive while
the $^3P_1$ wave shows a resonance-like behavior. In order to check
if this behavior corresponds to a true resonance or it is
simply an effect of the channel coupling, we disconnected
the $^5P_2$ wave. The resulting Argand diagrams show a weakly attractive
behavior, similar to that of the $0^+$ and $0^-$ channels.
This shows that indeed
the resonance-like behavior observed in Fig. 4 is due simply
to channel coupling.

As stated before, the possibility that a
resonance may exist in the $0^-$ channel has been
the subject of great discussion recently \cite{BIL1,GAR1,BIL2}.
A $0^-$ resonance in
the isospin-0 channel was proposed in Ref. \cite{BIL1}.
However, it was
pointed out in Ref. \cite{GAR1} that
most likely the resonance must have isospin two. The proposed resonance
\cite{BIL1,GAR1,BIL2} has a mass of 2065 MeV
and a very small width of
0.51 MeV. It is therefore very
interesting to investigate whether a model like the one used in
this work could
predict a resonance with the known features of this resonance, and
particularly with such a tiny width.

In order to do this, we have artificially varied the mass of the
$\sigma$ meson such as to increase the amount of attraction (the coupling
constant of the $\sigma$ is fixed by chiral symmetry). We found that if
$m_\sigma <$ 375 MeV a resonance appears in the $0^-$ channel. We
show in Table III the mass and width of this resonance for several values
of $m_\sigma$. As it can be observed from this table, the width of the
resonance drops dramatically when its mass approaches the
$\pi NN$ threshold (2017 MeV).
This result can be understood from simple angular momentum barrier
considerations. If we call $q$ and $L$ to the relative momentum and
relative orbital angular momentum between a nucleon and the $\pi$-nucleon
pair, respectively, then since $L=1$ the
width of the resonance will be proportional
to $q^{2L+1}=q^3$, so that it will drop very fast as one approaches the
$\pi NN$ threshold since there $q \to 0$.

As shown in Table III, when the mass of the sigma is taken as
$m_\sigma=$ 234 MeV the mass of the resonance has the
experimental value 2064.4 MeV and a
width of 0.6 MeV, which is in very good agreement with
the value of the width extracted by Bilger
and Clement \cite{BIL1}. This change in the mass of the sigma
generates additional attraction in the other channels, giving rise
in some cases to resonances. However, all of them
lie above the $0^-$ resonance, which shows that the $0^-$ channel
is the most attractive one. Therefore, the sharp peak seen in the
double charge exchange reactions could be justified as a
nucleon-$\Delta$ resonance in the isospin 2 channel, without
resorting to exotic decay channels.

We have generated artificially the additional attraction
needed to produce the $0^-$ resonance by varying the mass of the
sigma meson. In principle, however, a possible source of this lacking
of attraction within the quark cluster model could be the inclusion
of the non-local terms of the nucleon-$\Delta$ interaction
and the contribution of the $\Delta - \Delta$ channels. Both of
these effects have been neglected in the present study.

\section{Conclusions}

We have studied the nucleon-$\Delta$ system in the isospin-2 channels
within the quark cluster model of the baryon-baryon interaction. We
have found that this model with standard parameters derived from
nucleon-nucleon scattering is unable to generate any resonance
with $J \le 2$. However, we found that the $0^-$ channel is the most
attractive one. We have varied freely the mass of the sigma meson
in order to generate a resonance in this channel. When the mass of
the $0^-$ resonance reaches the value of 2065 MeV, its width is equal
to 0.6 MeV. This very narrow width is in very good
agreement with the width extracted in Refs. \cite{BIL1,GAR1,BIL2}.

\acknowledgements

H.G. thanks the Nuclear Physics Group at the University of
Salamanca for their kind hospitality.
This work has been partially funded by Direcci\'on
General de Investigaci\'on Cient\'{\i}fica y T\'ecnica
(DGICYT) under the Contract No. PB91-0119-C02 and by
COFAA-IPN (Mexico).

\begin{figure}
\caption{ Fredholm determinat of the nucleon-$\Delta$
channels with $J \le 2$ as a function of the non-relativistic
energy E.}
\label{fig1}
\end{figure}

\begin{figure}
\caption{ Phase shift of the $0^+$ nucleon-$\Delta$ channel as a function
of the invariant mass $W=S^{1/2}$ .}
\label{fig2}
\end{figure}

\begin{figure}
\caption{ Phase shift of the $0^-$ nucleon-$\Delta$ channel as a function
of the invariant mass $W=S^{1/2}$.}
\label{fig3}
\end{figure}

\begin{figure}
\caption{ Argand diagrams of the diagonal amplitudes of the
$^3P_2$, $^5P_2$ and
$^3F_2$ nucleon-$\Delta$ partial waves. The dots correspond
to the energies 2.12 to 2.44 GeV.}
\label{fig4}
\end{figure}

\begin{table}
\caption{ Quark model parameters.}
\label{quark}

\begin{tabular}{cccc}
 & $m_q (MeV)$                   &  313    & \\
 & $b (fm)$                      &  0.5    & \\
\tableline
 & $\alpha_s$                    &  0.4    & \\
 & $a_c (MeV \cdot fm^{-2})$     &  57.96  & \\
 & $\alpha_{ch}$                 &  0.0288 & \\
 & $m_\sigma (fm^{-1})$          &  3.42   & \\
 & $m_\pi (fm^{-1})$             &  0.7    & \\
 & $\Lambda (fm^{-1})$           &  4.2    & \\
\end{tabular}
\end{table}

\begin{table}
\caption{ Comparison of the nature of the various
$N\Delta$ channels with isospin
two to those predicted by three body models.}
\label{models}

\begin{tabular}{cccccc}
 & Channel &  Ref. [4] & Ref. [5] &  This work & \\
\tableline
 & $0^+$ & Repulsive       & Attractive      & Attractive     & \\
 & $0^-$ & Attractive      & Very attractive & Attractive     & \\
 & $1^+$ & Repulsive       & Repulsive       & Repulsive      & \\
 & $1^-$ & Repulsive       & Attractive      & Repulsive      & \\
 & $2^+$ & Repulsive       & Repulsive       & Very repulsive & \\
 & $2^-$ & Very attractive & Very attractive & Repulsive      & \\
\end{tabular}
\end{table}

\begin{table}
\caption{ Dependence of the mass and width of the $0^-$ resonance on
the mass of the sigma.}
\label{width}

\begin{tabular}{ccccc}
 & $m_\sigma (MeV)$       & $M_{Res} (MeV)$  &  $\Gamma_{Res} (MeV)$& \\
\tableline
 & 350.0                         &  2170.4      &  127.6         & \\
 & 300.0                         &  2128.7      &  18.2          & \\
 & 250.0                         &  2083.4      &  2.3           & \\
 & 234.0                         &  2064.4      &  0.6           & \\
 & 210.0                         &  2029.5      &  0.0015        & \\
\end{tabular}
\end{table}

\end{document}